\documentclass[multphys,vecphys]{svmult}

\usepackage{makeidx}         
\usepackage{graphicx}        
\usepackage{multicol}        
\usepackage[bottom]{footmisc}

\makeindex             

\begin{document}

\title*{Inspiral of double black holes in gaseous nuclear disks}

\author{Massimo Dotti\inst{1}\and Monica Colpi\inst{2}\and 
Francesco Haardt\inst{3}}
\institute{Universit\`a dell'Insubria, Como
\texttt{dotti@mib.infn.it}
\and Universit\`a di Milano-Bicocca, Milano
\texttt{colpi@mib.infn.it}
\and Universit\`a dell'Insubria, Como
\texttt{haardt@mib.infn.it}}
%
%
\maketitle

\begin{abstract}
We study the inspiral of double black holes
orbiting inside a massive  
rotationally supported gaseous disk, with masses in the 
Laser Interferometer Space Antenna ({\it LISA}) 
window of detectability.
Using high--resolution SPH 
simulations, we follow the black hole dynamics in the early phase 
when gas--dynamical friction acts on the black holes individually,
 and continue our simulation until the form a close binary.
We find that in the early sinking the black holes loose memory
of their initial orbital eccentricity if  they co--rotate with 
the gaseous disk. As a consequence the massive black holes 
{\it form a binary with very low eccentricity}.  
During the inspiral, gravitational capture of gas by 
the black holes occurs mainly when they move on circular orbits and may
ignite AGN activity: eccentric 
orbits imply instead high relative velocities and weak gravitational 
focusing. 
\end{abstract}

\section{Introduction}
\label{sec:1}

Close massive black hole (MBH) binaries are natural, 
powerful sources of gravitational
radiation, whose emission is one of the major scientific targets of
{\it LISA} (see, e.g.,
Haehnelt 1994, Jaffe \& Backer 2003, Sesana et al. 2005). 
How can MBHs reach sub--parsec distance scales and coalesce 
when resulting  from the collision and merger of galaxies?
Recently, Kazantzidis et al. (2005) explored the effect of 
gaseous dissipation in mergers between gas--rich disk galaxies 
with central MBHs, using high resolution N--Body/SPH simulations.  
They found that the interplay between strong gas 
inflows, cooling processes, and star formation, leads 
naturally to the formation of a close MBH pair and of
massive nuclear gaseous disks around, on a scale of $\sim 100$ pc,
close to the numerical resolution limit (updated 
simulations are presented by Mayer et al. in these proceedings). 
On smaller scales, 
Escala et al. (2005) have studied the role of gas on the 
orbital evolution of MBH binaries as a function of disk clumpiness,  
MBH to gas mass ratio, and orbital inclination 
angle.
In the same context, we studied the evolution of the eccentricity,
a key parameter for assessing the role played by gravitational wave 
emission in the orbital decay of MBH binaries in the {\it LISA} 
mass range.

\section{The simulations}
\label{sec:2}

We perform our simulations with MBHs embedded in 
a spheroidal component (bulge) modeled initially as a
Plummer sphere, and in a Mestel gaseous
disk.  
We evolve the system using the N--Body/SPH code GADGET 
(Springel, Yoshida \& White 2001).

We show the case of two MBHs of mass $10^6\, {\rm M_{\odot}}$ 
and $5\times 10^6\, {\rm M_{\odot}}$ orbiting in the disk 
plane. 
We allow the heavier MBH (MBH1) to reach the center before the 
sinking process of the light one (MBH2) takes place.
MBH2 is initially moving on a prograde eccentric orbit ($e=0.95$). 
Figure 1, upper panel, shows
the MBH distances from the center of mass, as a function of 
time. We find that
the sinking time of MBH2 is $\simeq 10^7$ yr, and that
there is a pronounced circularization of its orbit.  
Lower panel shows the gas mass collected 
by the  two MBHs   during  their  orbital  evolution.  Only  when 
the  orbit of MBH2 becomes circular  gathering of gas can occur.
Figure 2 shows, at two selected times, the face-on projection of
the gas density together with the orbit of MBH2:
close to pericentre the MBH has a speed larger, 
in modulus, than the local gas speed so that the wake is excited behind 
its trail, eroding the radial component of the velocity (left panel). On the  
other  hand,  near to apocentre, the MBH tangential velocity is 
slower than the local rotational gas velocity, so that the wake 
is dragged in front of the MBH, increasing its angular momentum (right panel).
The net effect is the circularization of the initially eccentric 
MBH orbit. New simulations with pure stellar disks present the same effect.

\begin{figure}
\centering
\includegraphics[height=6cm]{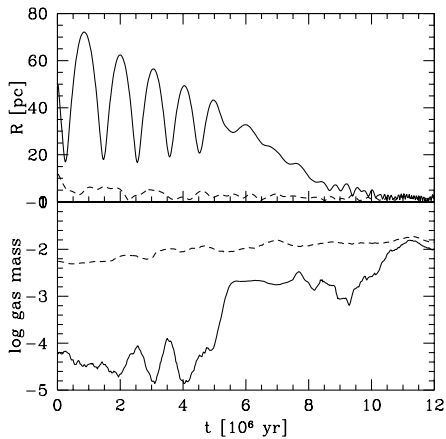}
\caption{Upper panel: Solid (dashed) line 
shows the distance R (pc) of MBH2 (MBH1) from 
the center of mass as a function of time. Lower panel: Solid 
(dashed)  line shows the mass of  the  over-density corresponding 
to MBH2 (MBH1) as a function of time.}
\label{fig:1}       
\end{figure}

\begin{figure}
\centering
\includegraphics[height=6cm]{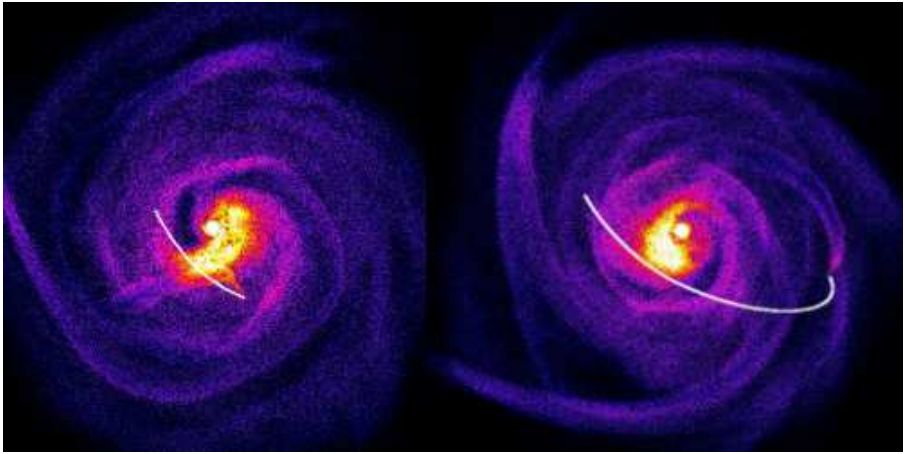}
\caption{Time sequence of the sinking of the MBH2.  The color 
coding indicates the z--averaged gas density (in logarithmic scale), 
and the white line traces the MBH2 counterclockwise prograde orbit.
In the left panel the
over--density created by  MBH2 is behind its current trail, while
in the right panel, the BH finds its own wake in front of its path. 
The wake is dragged by the faster rotation of the disk.} 
\label{fig:2}       
\end{figure}

\bigskip

\section{Conclusion}
\label{sec:2}

Our main findings are:
\begin{itemize}
\item dynamical friction due to the MBH--gas interaction is effective 
down to pc distance scales
\item the MBH--disk interaction circularizes prograde orbits
\item substantial gas mass can be gathered along circular MBH orbits:
binary coalescence could be accompanied by (double) AGN activity.
\end{itemize}
For full details and references please see: Dotti, Colpi, Haardt: 
MNRAS (in press, astro-ph/0509813)
%
%
%
%
%

%
%



\printindex
\end{document}